\documentclass[aps,prb,superscriptaddress,showpacs, preprint
]{revtex4}
\usepackage{graphicx}
\usepackage{epstopdf}
\usepackage{amssymb,amsmath,amsfonts}

\newcommand{\coso} {{\mbox{Cu${}_2$OSeO${}_3$}}}
\begin{document}
\title{Dissipation processes in the insulating skyrmion compound $\coso$}
\author{I. Levati\'c}
\affiliation{Institute of Physics, Bijeni\v cka 46, HR-10 000, Zagreb, Croatia}
\author{V. \v Surija}
\affiliation{Institute of Physics, Bijeni\v cka 46, HR-10 000, Zagreb, Croatia}
\author{H. Berger}
\affiliation{Institute of Condensed Matter Physics, EPFL, CH-1015 Lausanne, Switzerland}
\author{I. \v Zivkovi\'c}
\email{zivkovic@ifs.hr}
\affiliation{Institute of Physics, Bijeni\v cka 46, HR-10 000, Zagreb, Croatia}

\date{\today}

\begin{abstract}
We present a detailed study of the phase diagram surrounding the skyrmion lattice (SkL) phase of Cu$_2$OSe$_2$O$_3$ using high-precision magnetic ac susceptibility measurements. An extensive investigation of transition dynamics around the SkL phase using the imaginary component of the susceptibility revealed that at the conical-to-SkL transition a broad dissipation region exists with a complex frequency dependence. The analysis of the observed behavior within the SkL phase indicates a distribution of relaxation times intrinsically related to SkL. At the SkL-to-paramagnet transition a narrow first-order peak is found that exhibits a strong frequency and magnetic field dependence. Surprisingly, very similar dependence has been discovered for the first-order transition below the SkL phase, i.e.~where the system enters the helical and conical state(s), indicating similar processes across the order-disorder transition.
\end{abstract}

\pacs{75.30.Kz, 75.40.Gb, 75.75.Jn}

\maketitle

%
%
%
%

\section{Introduction}
\label{Introduction}
As a part of the fast growing area of interes of both fundamental and applied science, spin textures in general and in particular skyrmions themselves have been in the focus of numerous studies in recent years. Magnetic skyrmions are defined as topologically protected field configurations with particle-like properties predicted to exist in chiral magnets~\cite{Rossler2006}. Described as vortex like spin textures of nanometer dimensions, they form a 2D hexagonal lattice under the influence of an applied external magnetic field. The plane of the lattice lies perpendicular to the direction of the magnetic field with the magnetization at the center of every such vortex, i.e. skyrmion, being antiparallel to the direction of the magnetic field. Such structures were found to exist in a small portion of the $H-T$ phase diagram of B20 metallic alloys such as MnSi~\cite{Muhlbauer2009}, FeGe~\cite{Yu2011} and Fe$_{1-x}$Co$_x$Si~\cite{Munzer2010, Yu2010}. Although the crystallized form of skyrmions seemed to be confined to this family of compounds, recent studies on an insulator compound Cu$_2$OSe$_2$O$_3$ have shown that as well as sharing the $P2_13$ crystal structure of the B20 alloys, analogous to 
theirs, its phase diagram also contains a skyrmion lattice (SkL) phase. The existence of skyrmions in Cu$_2$OSe$_2$O$_3$ was confirmed in both thin film and bulk samples using Lorentz transmission electron microscopy (Lorentz TEM)~\cite{Seki2012} and small angle neutron scattering (SANS)~\cite{Adams2012} respectively.

In addition to being interesting from purely the perspective of being the only insulator found so far to host a skyrmion lattice phase in its phase diagram, Cu$_2$OSe$_2$O$_3$ also shows promise when it comes to potential uses as several recent studies have shown some extent of success in controling the skyrmions in this compound. Firstly, a rotation of the whole skyrmion lattice was achieved by applying an electric field while crossing into the skyrmion phase~\cite{White2012}. Secondly, three different resonant modes (clockwise, counterclockwise and breathing mode) have been detected with oscillating magnetic fields at frequencies of $1$ to $2\:$GHz~\cite{Okamura2014}.

The ground state of Cu$_2$OSe$_2$O$_3$ is a helimagnet~\cite{Adams2012}, similar to other SkL-compounds. It arises from the competition of energy scales of various ordering mechanisms. While the energetically favoured ferromagnetic exchange prefers a parallel spin alignment, in the case of noncentrosymmetric systems such as the $P2_13$ structures, a modulation of the spin alignment by the weaker Dzyaloshinskii-Moriya interaction preferring a perpendicular spin orientation produces a long wavelength helical spin ordering. When the magnetic field is applied, the system transforms into a conical state. SkL is formed for a moderate magnetic field close to the ordering temperature and it can be described as a triple-helix structure perpendicular to the direction of the magnetic field~\cite{Muhlbauer2009}.

One of the standing issues related to SkL is how it forms out of the paramagnetic state and also what is the nature of the transition from SkL into the conical state at lower temperatures. Recent reports have provided contradictory claims: on the one side the formation of SkL has been explained in terms of a single thermodynamic phase, driven by thermal Gaussian fluctuations~\cite{Muhlbauer2009, Bauer2013}, with experiments performed on MnSi. On the other side there is a prediction of various topological textures governed by magnetic anisotropy~\cite{Butenko2010} and experimentally supported by measurements on FeGe~\cite{Wilhelm2011}. To the best of our knowledge, there have been no reports on Cu$_2$OSe$_2$O$_3$ addressing those questions. In this paper we provide a detailed study of the dynamics of the SkL and transitions into surrounding phases by measuring the magnetic susceptibility of the system. The imaginary component of the susceptibility, which is related to dissipation processes occurring within one cycle of the alternating magnetic field, reveals a narrow peak associated with the first-order phase transition between the SkL state and the paramagnetic state. At the same time, the transition to the conical state is characterized with a relatively broad dissipation region indicating different type of processes governing the disintegration of SkL.

The paper is organized as follows: in section \ref{Details} we provide the details of the experiment. In Section \ref{Results} we present the experimental results which are analyzed and discussed in section \ref{Discussion}. Finally, in section \ref{Conclusion} we make a conclusion.

%
%
%
%

\section{Experimental details}
\label{Details}
The single crystal sample (synthesis details given in~\cite{Zivkovic2012}) used in our measurements was elongated along the $\left[111\right]$ direction with dimensions $4 \times 1 \times 1\:$mm$^3$. We have aligned the sample parallel to the direction of the external magnetic field which was supplied using a commercial $9\:$T magnet. Additional fine tuning of the applied magnetic field was done using the primary coils of the home-made susceptometer and a direct current source (Keithley6221). Measurements were done by performing temperature scans at applied field steps of $5\:$Oe or less from zero field up to $340\:$Oe. All the measurements were taken in a cooling regime ($5\:$mK/s) at a given fixed external field value starting from $60\:$K, well above the ordering temperature. The value of the external field was changed only after all the measurements at various frequencies have been performed. The excitation field was $1\:$Oe at all frequencies.

%
%
%
%

\section{Results}
\label{Results}

\begin{figure}
\includegraphics[width=0.35\textwidth]{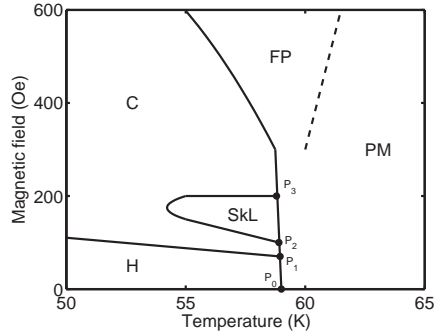}
\caption{A sketch of the phase diagram for $\coso$. Similar diagrams are found in all SkL-compounds.}
\label{fig-sketch}
\end{figure}

A well established phase diagram around the SkL pocket is schematically presented in Fig.\ref{fig-sketch}. At low fields the helical ground state occurs below the transition temperature $T_N$ ($P_0$ in Fig.~\ref{fig-sketch}). As the field increases the moments form a conical arrangement around the direction of the magnetic field. At still larger fields the moments become disordered but field-polarized. The dashed line extending diagonally into the paramagnetic regime represents a cross-over from such a field-polarized region and a fully disordered paramagnetic state at high temperatures which has been recently shown to follow a classical 3D Heisenberg scaling~\cite{Zivkovic2014}. The SkL pocket is located within the conical phase, spanning $2-3\:$K just below the order-disorder line between points $P_2$ and $P_3$.

%
%

Such a phase diagram has been established using various techniques, microscopic using neutrons (small angle neutron spectroscopy, SANS)~\cite{Adams2012} but also bulk methods such as magnetization~\cite{Adams2012}, polarization~\cite{Seki2012a} and magneto-electric susceptibility~\cite{Omrani2014}. Often, due to small changes observed when crossing the boundaries, most notably between the conical and the SkL phases, a derivative of the magnetization dM/dB is calculated to mark the boundary. A direct measurement of the ac magnetic susceptibility $\chi$ is favoured in such cases for several reasons: (i) it avoids the errors and averaging of the numerical approach; (ii) it allows to investigate the time-domain by changing the frequency of the excitation field; (iii) it provides another quantity, the out-of-phase component $\chi ''$ or the imaginary part of the total susceptibility $\chi = \chi ' + i\chi '' = \operatorname{Re}\chi + i\operatorname{Im}\chi$. $\operatorname{Im}\chi$ serves as a measure of the dissipation in a magnetic system within one cycle of the ac field and is related to the size of the hysteresis loop in the M -- H plane. It is often used in systems where a distribution of relaxation times is found, such as superparamagnets and spin-glasses~\cite{Mydosh1993}.

\begin{figure*}
\includegraphics[width=0.9\textwidth]{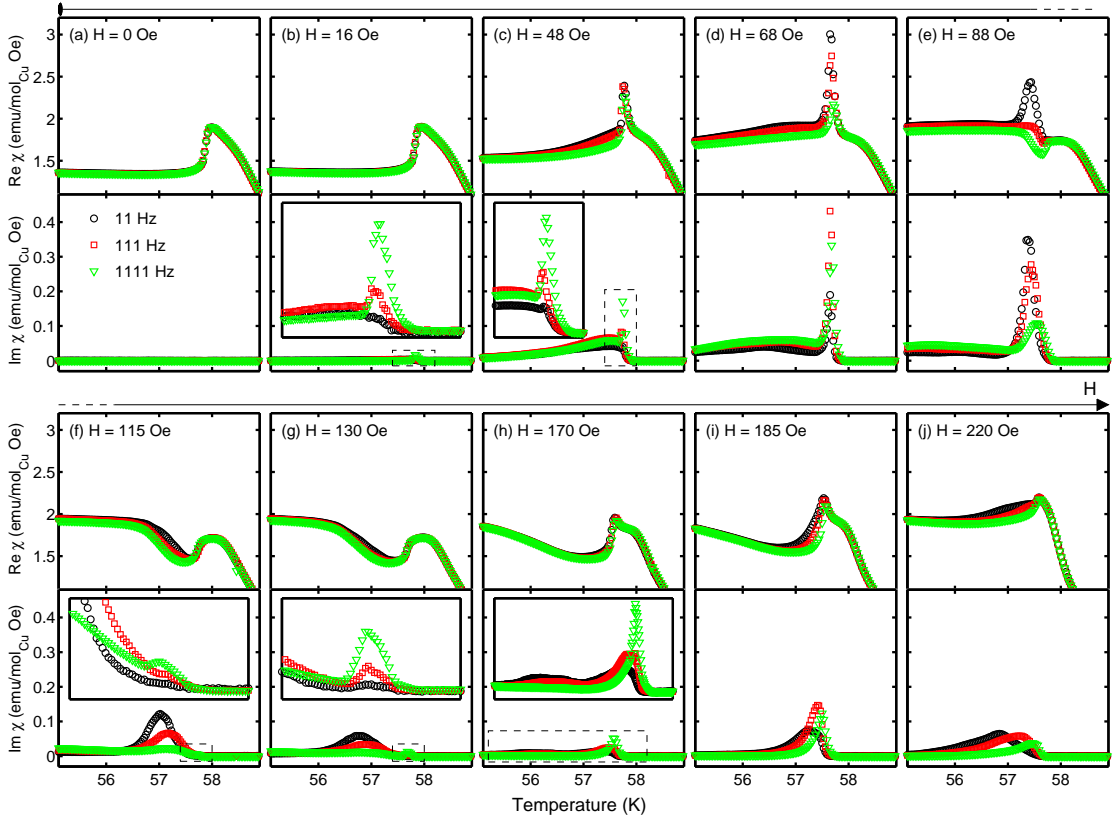}
\caption{(Color online) The evolution of the real and the imaginary part with magnetic field across the order-disorder transition. Insets show a magnified view of the region within dashed rectangles.}
\label{fig-evolution}
\end{figure*}

At zero dc magnetic field $\operatorname{Re}\chi$ exhibits a well-known behavior as previously published, Fig.~\ref{fig-evolution}(a). In the paramagnetic regime the susceptibility rises as temperature is lowered until around $59\:$K where an inflection point exists~\cite{Zivkovic2012}. A first order phase transition is marked as a sharp drop in $\operatorname{Re}\chi$ around $58\:$K below which a shallow minimum follows. Within our detection limit, $\operatorname{Im}\chi = 0$ in the whole range of investigated temperatures.

%
%

The situation changes when a dc magnetic field is applied. In Fig.\ref{fig-evolution}(b) we show the transition at $H = 16\:$Oe. The real part has remained similar to $H = 0$, however $\operatorname{Im}\chi$ develops a sharp peak at $1111\:$Hz with a kink at the low-$T$ part which marks the entrance to the helical phase (see the insets for magnified views). The peak is also visible at $111\:$Hz but much smaller in amplitude. At $11\:$Hz only a broad feature is seen which has a maximum exactly at the position of the kink at higher frequencies. Only when the field is further increased does the $11\:$Hz curve develop a small peak as seen in Fig.\ref{fig-evolution}(c) for $H = 48\:$Oe. At the same time the level of overall dissipation at the transition and in the ordered region has grown by one order of magnitude. Also, a second, broad maximum can be seen for all the curves that represents the transition between the helical and conical phases. As for $\operatorname{Re}\chi$, instead of a drop in the susceptibility the region around the transition has developed a sharp peak.

At $H = 68\:$Oe, Fig.\ref{fig-evolution}(d), we notice that the maximum dissipation at the transition is no longer achieved with $1111\:$Hz but with $111\:$Hz, while $11\:$Hz takes over at $H = 88\:$Oe (Fig.\ref{fig-evolution}(e)). At the same time the peaks start to broaden and $\operatorname{Re}\chi$ at $1111\:$Hz develops a minimum familiar from previous studies as a hallmark of the SkL phase. At lower frequencies the minimum develops at somewhat larger fields where the dissipation has been shifted to lower temperatures and the width continues to increase.

At $H = 115\:$Oe a second peak in $\operatorname{Im}\chi$ develops around $57.7\:$K, Fig.\ref{fig-evolution}(f), which marks the transition between the SkL and the paramagnetic phase. The SkL phase is now fully developed as can be seen by a pronounced minimum in $\operatorname{Re}\chi$ for all the frequencies. Again, the most prominent dissipation at the transition from the SkL to paramagnetic phase is visible with $1111\:$Hz while the susceptibility measured with $11\:$Hz develops a peak at larger fields. The real part also begins to develop a peak on top of a rounded feature at $58\:$K, similar to the development of a peak at lower fields, Fig.\ref{fig-evolution}(c). The real part at $H = 115\:$Oe and $130\:$Oe also exhibits a drop, although not as sharp as the one seen for $H = 0$.

Approaching the $P_3$ point we can recognize three different regions of dissipation, especially well pronounced at $111\:$Hz, Fig.\ref{fig-evolution}(h): the transition from the paramagnetic phase is marked with a sharp peak, then it is followed by a broad feature and then below a shallow minimum another broad dissipation region, the continuation of features seen at lower fields. Similar to the region between $P_0$ and $P_2$, at some values of the magnetic field the dissipation at $111\:$Hz and $11\:$Hz becomes the most pronounced, Fig.\ref{fig-evolution}(i) and Fig.\ref{fig-evolution}(j), respectively. Above $H = 220\:$Oe the dissipation profiles remain practically the same, with a diminishing amplitude until it goes below the detection limit of our setup.

%
\begin{figure}
\includegraphics[width=0.45\textwidth]{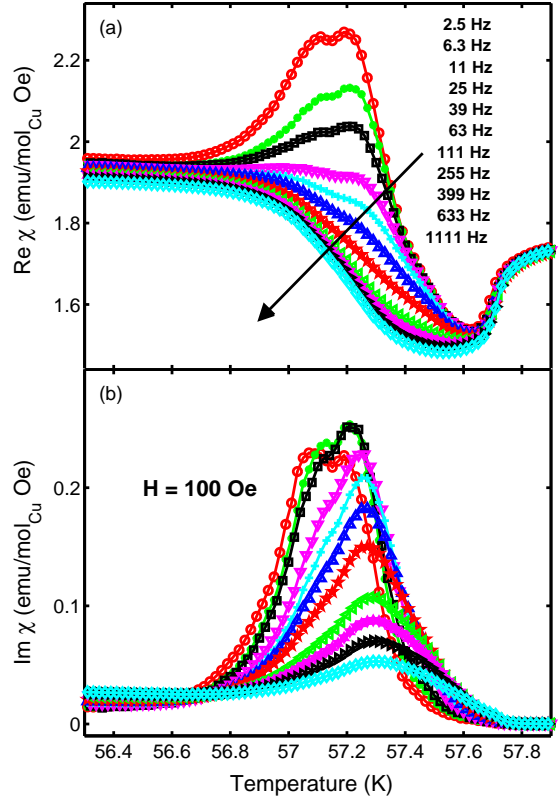}
\caption{(Color online) (a) The real and (b) the imaginary part of the susceptibility at $H = 100$ Oe.}
\label{fig-11freqs100Oe}
\end{figure}
%

\begin{figure}
\includegraphics[width=0.45\textwidth]{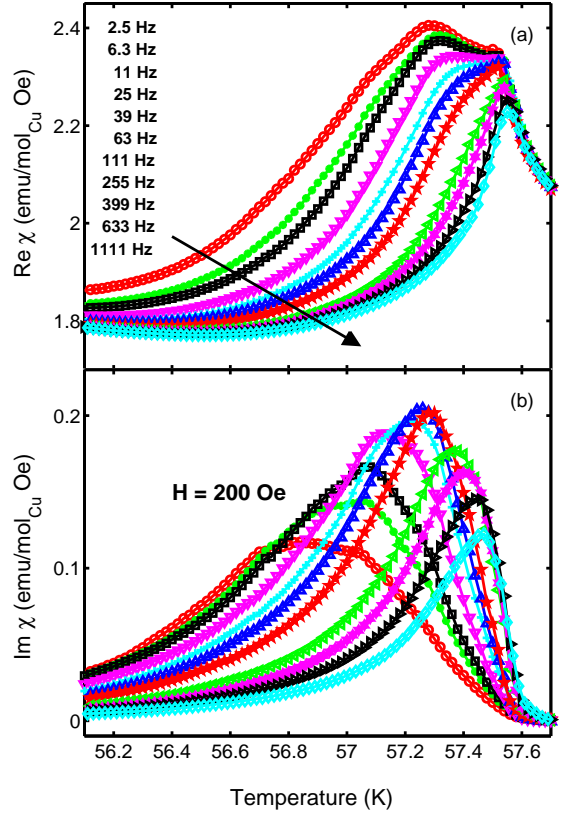}
\caption{(Color online) (a) The real and (b) the imaginary part of the susceptibility at $H = 200$ Oe.}
\label{fig-11freqs200Oe}
\end{figure}

In order to investigate the transition between the conical and SkL phases more closely, we have performed a systematic investigation of the frequency dependence at various magnetic fields. In Fig.\ref{fig-11freqs100Oe} we present temperature scans at frequencies $2.5\:$Hz, $6.3\:$Hz, $11\:$Hz, $25\:$Hz, $39\:$Hz, $63\:$Hz, $111\:$Hz, $255\:$Hz, $399\:$Hz, $633\:$Hz and $1111\:$Hz for $H = 100\:$Oe that characterizes the lower-left conical-SkL boundary, and in Fig.\ref{fig-11freqs200Oe} for $H = 200\:$Oe that characterizes the upper boundary of the SkL phase. As seen already in Fig.\ref{fig-evolution}, the lower-left boundary shows a broadened dissipation region compared to a narrow peak at the main transition, with a valley in the real part due to the formation of SkL. However, at low frequencies the real part develops a maximum with two well-defined peaks $\approx 150$ mK apart. The double-peak feature can also be recognized in the imaginary part, where the lower peak quickly diminishes as the frequency is increased. Since this is the first observation of this feature, we can only speculate about its origin. A viable candidate might be anisotropy. Namely, in the SkL phase the triple-helix lies in the plane perpendicular to the direction of the magnetic field. In the process of transforming this structure into a conical phase, there might be a (metastable) phase where the direction of helices is given by magneto-crystalline anisotropy ([100] for $\coso$). Also, several topological phases have been recently suggested which might explain the double-peak feature~\cite{Wilhelm2011}.

It is interesting to notice how at high frequencies the dissipation starts quite sharply (the kink below $57.8\:$K) while at lower frequencies the emergence of dissipation is relatively smooth. On the other side, around $56.7\:$K, an unusual inversion occurs. Namely, just above that temperature the dissipation grows as the frequency is decreased. Below, the pattern is inverted, with dissipation growing at higher frequencies. Around that temperature the double-peak structure in the real part emerges from the monotonic low temperature background, indicating that this is where the formation of larger spin structures begins, that ultimately lead to skyrmions.

%
%

At higher magnetic fields qualitatively no significant changes occur. The imaginary component decreases and around $150\:$Oe it is practically zero for all the frequencies within our experimental range (not shown).

%
\begin{figure*}
\includegraphics[width=0.9\textwidth]{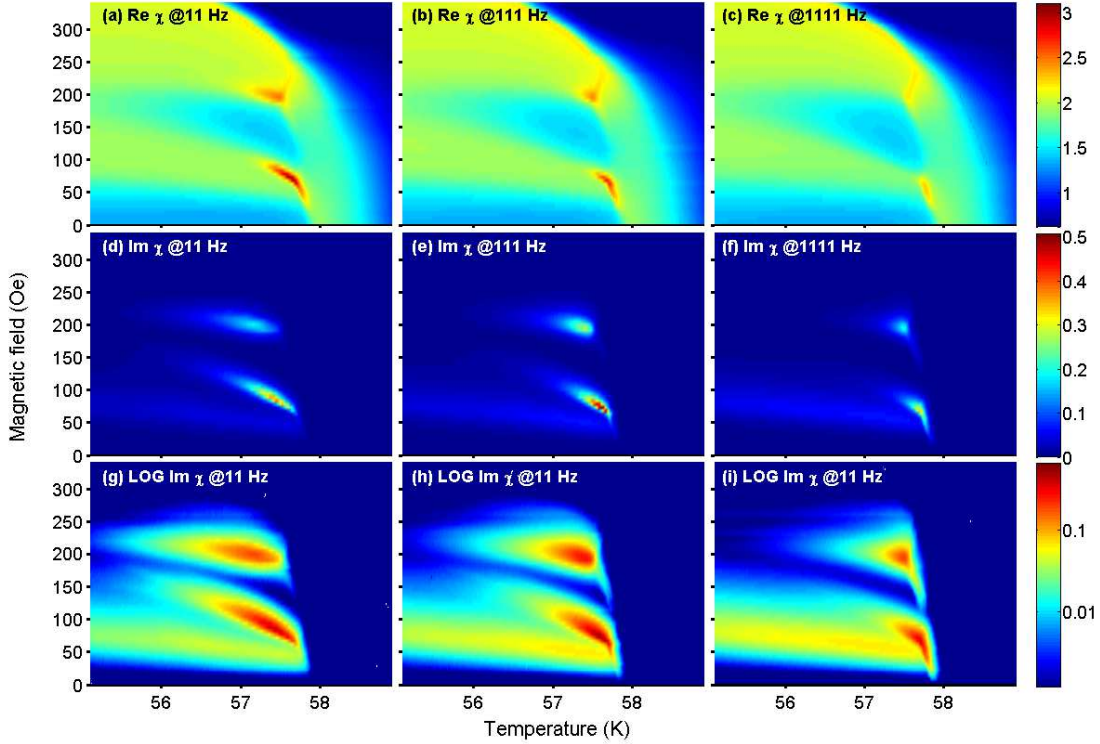}
\caption{(Color online) Color-plots of the real and the imaginary part of the susceptibility at three different frequencies. The units in color-bars are given in emu/mol$_{Cu}$ Oe.}
\label{fig-maps}
\end{figure*}
%

The frequency dependence for the upper boundary of the SkL phase exhibits prominent differences compared to the lower-left boundary, see Fig.~\ref{fig-11freqs200Oe}. At high frequencies the real part exhibits only one peak around $57.5\:$K that marks the transition from the paramagnetic state. However, lowering the frequency reveals another peak at $\approx 57.2\:$K for $f = 2.5\:$Hz, similar to the observed appearance of a double-peak structure in Fig.~\ref{fig-11freqs100Oe}. On the other hand, the imaginary component exhibits only one peak with a significant frequency dependence. The dissipation starts around the same temperature and then for high frequencies it sharply rises to form a peak with an exponential tail while for the lower frequencies it forms a broad maximum with several shoulders, i.e.~it seems that the dissipation profile is structured.

%
%
%
%

\section{Analysis and Discussion}
\label{Discussion}

We attempt now to provide a more comprehensive view of the measured data. With increasing magnetic field, one can follow various trends in Fig.\ref{fig-evolution}, with maximums and minimums appearing, shifting and disappearing. However, the information about the overall picture is better depicted by plotting all the curves in one plot for three different frequencies, $11\:$Hz, $111\:$Hz and $1111\:$Hz, that have been used to map the entire range of magnetic fields from $0-340\:$Oe. In Fig.\ref{fig-maps} we show the real and the imaginary component of the susceptibility, together with the imaginary component in the logarithmic scale, in order to emphasize the low-lying features.

The color-plots of the real part (Fig.\ref{fig-maps}(a)-(c)) reveal a familiar shape of the SkL phase embedded within the conical phase below which a ground-state helical phase is located. The boundary between ordered states and the disordered state that starts at $T_N = 57.9\:$K for $H = 0$ shifts slightly to lower temperatures at higher fields. Surprisingly, above the $P_3$ point the boundary bends back towards higher temperatures before another change of the direction after which it continues towards $T = 0\:$. Such a behavior has not been reported so far for any of the known SkL-compounds.

The largest dissipation occurs along the boundaries between the conical and the SkL phase but it is clearly much more emphasized on the high temperature side of the boundaries (Fig.\ref{fig-maps}(d)-(f)). A clear frequency dependence can be observed in terms of the length of the dissipation region: at $11\:$Hz the dissipation extends approximately down to the half of the SkL pocket while at $1111\:$Hz the dissipation is visible only close to the order-disorder boundary. Similar observation of the frequency dependence can be noticed in the real part where the maximums around the $P_2$ and $P_3$ points are much stronger at lower frequencies. Additionally, in Fig.~\ref{fig-maps}(f) a narrow tail of dissipation can be noticed below the $P_2$ point which marks the order-disorder boundary (representative temperature scans are shown in Fig.~\ref{fig-evolution}(b) and (c)).

When plotting the imaginary part on the logarithmic scale, Fig.\ref{fig-maps}(g)-(i), some interesting low-lying features emerge. The conical-to-SkL regions are now extended towards lower temperatures but without an overlap. The dissipation associated with the helical-to-conical transition becomes clearly visible, with a weak frequency dependence. Finally, a narrow dissipation profile marks the SkL-to-paramagnet boundary, which has been noticed as a second peak in Fig.~\ref{fig-evolution}(f) and (g).

Here we should emphasize that from our bulk measurements we do not see any indication of the reported sub-structure of the skyrmion pocket~\cite{Seki2012}. To be completely correct, the two orientations of the skyrmion lattice suggested by Seki et al. should have an identical response to an alternating magnetic field applied along the length of skyrmions, as in our case. However, we would expect some level of dissipation at the boundary between the two orientations because of the formation of irregular spin clusters while the lattice adopts a new orientation. As can be seen from Fig.~\ref{fig-maps}, our high-sensitivity measurements do not reveal any trace of a sub-structure within the skyrmion pocket, not even in the logarithmic scale. This result is corroborated by a recent thorough mapping of the skyrmion pocket by SANS~\cite{White2014}.

\begin{figure}
\includegraphics[width=0.45\textwidth]{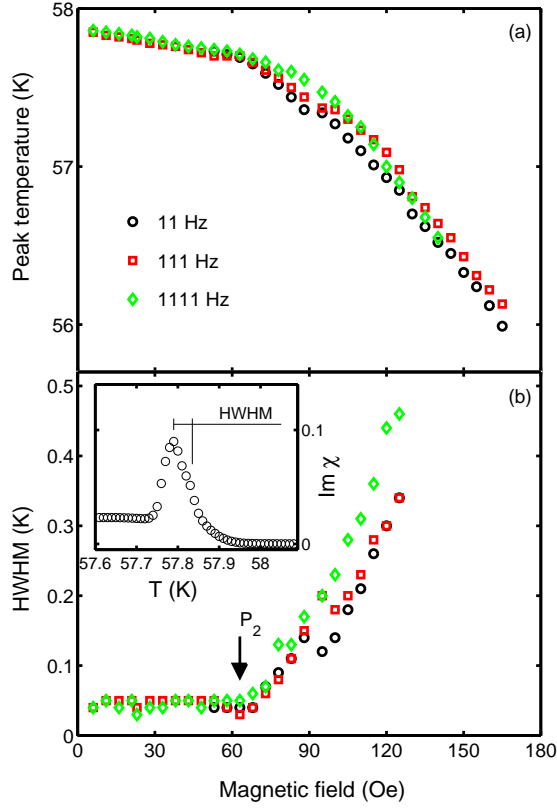}
\caption{(Color online) The field dependence of (a) the peak temperature of the imaginary part of the susceptibility and (b) half-width-at-half-maximum (HFHM). INSET: Temperature dependence of the imaginary part at $H = 33\:$Oe measured with $f = 1111\:$Hz. HWHM is measured on the high-temperature side of the peak.}
\label{fig-peakwidth}
\end{figure}

A careful inspection of Fig.\ref{fig-maps} reveals that the size of the SkL pocket as revealed from the real part of the susceptibility has a weak dependence on the frequency of the driving field, dominantly around $P_2$. The lower right corner, which is commonly associated with the $P_2$ point, is extended towards lower fields as the frequency increases. This has already been noticed in Fig.\ref{fig-evolution} where the characteristic valley shows up at $1111\:$Hz, while for lower frequencies the magnetic field has to be increased further for the same effect. A better method to determine $P_2$ is to use features observed in $\operatorname{Im}\chi$. As can be seen from Fig.\ref{fig-maps} the lower-left boundary of the SkL phase is characterized with a strong and relatively broad dissipation region. Plotting the position of the peak against the magnetic field for each frequency reveals that around $60\:$Oe there is a kink and curves begin to split, see Fig.\ref{fig-peakwidth}(a). Also, the half-width of the peak (Fig.\ref{fig-peakwidth}(b)) remains constant below $60\:$Oe while above it starts to spread out, indicating the start of the lower-left conical-SkL boundary. This leads us to conclude that $P_2$ is well defined and that for our sample its position is at $H(P_2) = 63\:$Oe and $T(P_2) = 57.7\:$K.

%
%

The determination of $P_3$, on the other hand, is straight-forward, since the upper conical-to-SkL boundary is practically flat. Although at $11\:$Hz the dissipation is extended towards lower temperatures, at $1111\:$Hz a well-defined maximum can be found at $H(P_3) = 195\:$Oe and $T(P_3) = 57.5\:$K.

Although both dissipation regions at the boundary between the conical and the SkL phase extend towards lower temperatures, we do not observe their overlap, which is expected to occur at much lower frequencies than $11\:$Hz. Overall, such a behavior is similar to a classical super-paramagnetic system where lowering the temperature prevents short-range coupled magnetic clusters to follow the driving field and leaves them effectively frozen. We can exclude domain wall dynamics since there are no domains with opposite net magnetic moments neither in the conical phase nor in the SkL phase. Recent magnetic-force microscopy study supported by numerical simulations on Fe$_{0.5}$Co$_{0.5}$Si\cite{Milde2013} established that the transition of the SkL phase into the conical phase is accompanied with the formation of singular magnetic point defects, magnetic monopoles and anti-monopoles. They represent points where two individual skyrmions merge. In that context we can explain a finite $\operatorname{Im}\chi$ as a response of such defects to the oscillating magnetic field when they travel along the skyrmion direction and in the process jump over energy barriers.

%
\begin{figure}
\includegraphics[width=0.45\textwidth]{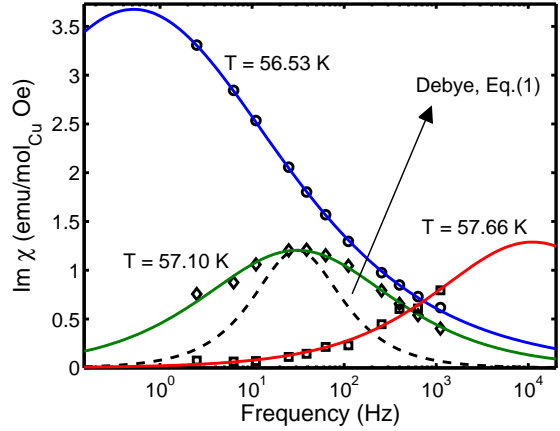}
\caption{(Color online) The frequency dependence of the imaginary part of the susceptibility at three different temperatures for $H = 130\:$Oe. The solid lines are fits using Eq.(\ref{EQimag}) for the distribution of relaxation times. The dashed line is a single $\tau_0$ model.}
\label{fig-fits}
\end{figure}
%

%
\begin{figure*}
\includegraphics[width=0.9\textwidth]{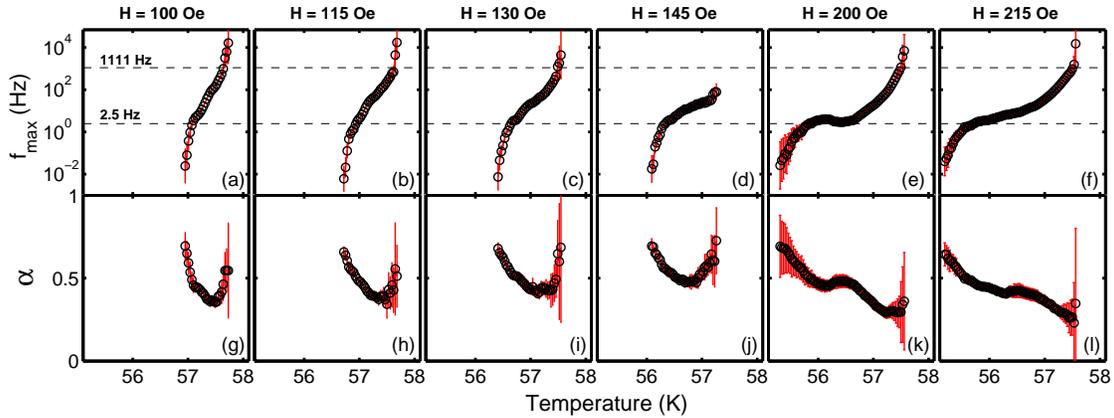}
\caption{(Color online) The temperature dependence of (a) the characteristic frequency $f_{max}$ and (b) the distribution width $\alpha$ extracted using Eq.(\ref{EQimag}) for several values of magnetic field. The error-bars are determined by the fit. The horizontal dashed lines represent the frequency range used in the experiment.}
\label{fig-parameters}
\end{figure*}
%

To investigate in more detail the dynamics and the nature of the SkL disintegration process we can examine the frequency dependence inside the SkL pocket. For that purpose we extract the values of the susceptibility from temperature scans measured at various frequencies (Fig.~\ref{fig-11freqs100Oe} and Fig.~\ref{fig-11freqs200Oe}) for a given magnetic field at equally spaced intervals (10 mK). As an example, in Fig.~\ref{fig-fits} we present three different cases obtained for $H = 130\:$Oe. Close to the conical-to-SkL border the maximum in the frequency response is achieved below the lowest frequency in our experiment ($2.5\:$Hz) while close to the SkL-to-paramagnet border the characteristic frequency is above the highest frequency ($1111\:$Hz). Consequently, one can find a narrow temperature window where the characteristic frequency is around $50\:$Hz, in the middle of our frequency range. In that case the points form a symmetric, bell-shaped curve. The maximum of such a curve is characterized by a condition $f(\operatorname{Im}\chi = \it{max}) = f_{max} = 1/(2 \pi \tau_0)$, where $\tau_0$ is the characteristic spin-relaxation time. For a system with a single $\tau_0$ the frequency dependence of the susceptibility follows a Debye model
\begin{equation}
\label{EQsingle}
\chi (\omega) = \frac{\chi (0)}{1 + i\omega \tau_0}
\end{equation}
and the imaginary component is represented in Fig.~\ref{fig-fits} with a dashed line. It is obvious that experimental points define a much wider profile implying that the system exhibits a distribution of relaxation times. In that case one can use\cite{Huser1986} a modified Eq.(\ref{EQsingle})
\begin{equation}
\label{EQdistribution}
\chi (\omega) = \frac{\chi (0)}{1 + (i\omega \tau_0)^{1-\alpha}}
\end{equation}
where $\alpha = 0$ reverts back to Eq.(\ref{EQsingle}) with a single relaxation time and $\alpha = 1$ gives an infinite width of the distribution. With $\alpha > 0$ $\tau_0$ becomes an average relaxation time. The frequency dependence of the imaginary component is given in that case by
\begin{equation}
\label{EQimag}
Im \; \chi (\omega) = \frac{1}{2} \chi_0 \frac{\cos \alpha \pi /2}{\cosh [(1-\alpha)\ln(\omega \tau_0)] + \sin \alpha \pi /2}
\end{equation}

Eq.(\ref{EQimag}) can be used to fit the data and extract the values of $f_{max}$ and $\alpha$ for a given temperature and magnetic field. Examples of the fit results are presented in Fig.~\ref{fig-fits} as solid lines.

One can then follow the evolution of parameters determined from the fit as we change temperature and magnetic field. We find that Eq.(\ref{EQimag}) is adequate in describing the behavior of the frequency profile in a wide temperature range, even when $f_{max}$ is outside our experimental window. In Fig.(\ref{fig-parameters}) we present the temperature dependence of $f_{max}$ and $\alpha$ for six fields that cut through the SkL pocket.

The temperature dependence of $f_{max}$ does not change qualitatively across the height of the SkL pocket. A quasi-linear dependence of $\log (f_{max})$ on temperature indicates an Arrhenius-like regime where a characteristic frequency depends on the height of the energy barrier $f_{max}(T) = f_0 exp(-\Delta /T)$. However, given that $f_{max}$ spans several orders of magnitude in a narrrow temperature range and that close to the SkL-to-paramagnet transition the curvature is opposite of that predicted by the classical Arrhenius law, we conclude that the energy barrier is also temperature dependent, $\Delta (T)$, and that it is decreasing as the temperature increases. This strongly suggests that the barriers are related to the stiffness of SkL and not to magnetic impurities, for which a characteristic frequency is expected to remain practically constant in such a narrow temperature range of $1-2\:$K where the SkL phase is found.

The width of the distribution of relaxation times remains in the range 0.3-0.5, except when close to transitions where it starts to increase. In that case $f_{max}$ is also drastically changed, especially at low frequencies where we see a freezing-like behavior. However, the error from the fit increases substantially so we cannot claim that the observed tendencies are intrinsic to the system and not just a consequence of $f_{max}$ being well away from our frequency range. Additionally, one needs to take into account the existence of a second peak that appears at frequencies below $10\:$Hz. Detailed experiments at lower frequencies and/or the relaxation of the magnetization could reveal the intrinsic behavior of the system close to the conical-SkL transition.

We now turn our attention to the SkL-to-paramagnet boundary. This boundary is characterized with a narrow, first-order peak. The peak is substantially larger at $1111\:$Hz compared to $11\:$Hz at all fields, indicating that small magnetic clusters are predominant within the transition, for which the characteristic frequency lies well above our range of frequencies. The fact that the amplitude grows going from $P_2$ to $P_3$ implies that larger structures are stabilized for which the characteristic frequency is smaller at larger fields (but still larger than $1111\:$Hz). It is tempting to identify those structures with field-stabilized individual skyrmions that form out of the paramagnetic state. There have been several different approaches in describing the system just above the ordering temperature. An extensive high-resolution neutron scattering study on MnSi by Pappas et al.\cite{Pappas2009} has revealed evidence for a chiral skyrmion spin liquid phase. In that context the freezing of individual skyrmions would be a natural precursor to the development of a long-range SkL. On the other hand, a different approach\cite{Janoschek2013} is based on the Brazovskii scenario, where strongly interacting chiral fluctuations are the main driving mechanism behind the fluctuation-induced first-order phase transition. It is important to notice that both studies have used SANS with $H = 0$ and that the exact development of fluctuations with magnetic field remains an open question. Below the transition the magnetic field successively transforms
%
\begin{figure}
\includegraphics[width=0.45\textwidth]{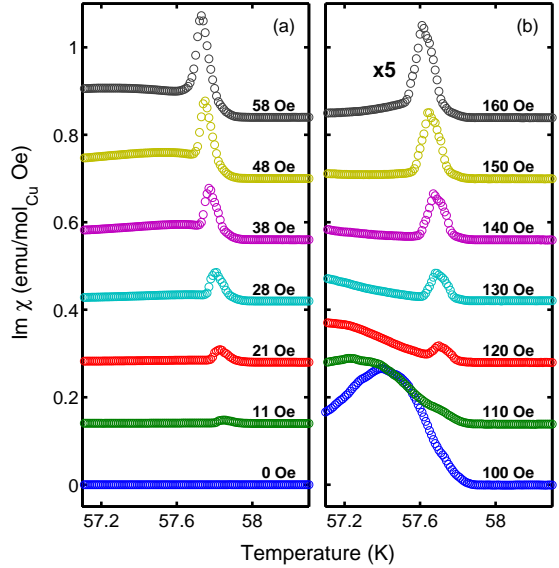}
\caption{(Color online) Temperature dependence of the imaginary part of the susceptibility measured with $f = 1111\:$Hz at several magnetic fields in the region (a) $P_0 \rightarrow P_2$ and (b) $P_2 \rightarrow P_3$. The curves have been shifted for clarity. The values in (b) are multiplied by factor 5.}
\label{fig-comparison}
\end{figure}
%
helical-to-conical-to-SkL-to-conical phases, however it is not clear whether the fluctuations above the transition change their character in the same manner and also how the transition itself is influenced. The recent specific heat study on MnSi\cite{Bauer2013} followed the evolution of a peak at the order-disorder transition with the magnetic field and concluded that gradually it weakens and transforms into the second-order peak, with a possible tricritical point well above the skyrmion pocket. The results from our experiments reveal a somewhat more complex behavior. The first-order peak in the imaginary component between $P_0$ and $P_2$ grows in amplitude and instead of continuing towards $P_3$, it widens and continues along the lower-left conical-to-SkL boundary. Between $P_2$ and $P_3$ a new peak emerges, again with growing amplitude, until it develops into a wide dissipation area around $P_3$. The evolution of the peak with the magnetic field for $P_0 \rightarrow P_2$ and $P_2 \rightarrow P_3$ for $f = 1111\:$Hz is presented in Fig.~\ref{fig-comparison} side-by-side. It is striking how similar the two paths look, although different processes within the first-order transition are expected to generate the dissipation with a different influence from the external $H$. Following the analogy between the two paths, one can infer from Fig.~\ref{fig-comparison} that a long-range SkL forms around $100\:$Oe, substantially higher compared to the position of the $P_2$ point ($63\:$Oe). This indicates that the region of the phase diagram surrounding the $P_2$ point represents a complex interplay between competing phases. The same could be implicated for the region around the $P_3$ point given the emergence of the second peak at low frequencies.

Finally, we would like to point out that the peaks observed at the order-disorder transition between $P_0$ and $P_2$ do not reflect the change of phases below the transition, i.e.~we do not 'see' the $P_1$ point. The weak dissipation profile between the helical and the conical phases is 'touching' the strong, narrow dissipation line of the first-order transition but it seems that there is no influence on the transition itself which appears smooth across the expected $P_1$ (see the log scale in Fig.\ref{fig-maps}). If the transition does not exhibit a qualitative change at $P_1$ then potentially the character of fluctuations above the transition could also be unaffected by the magnetic field. We hope that this finding will stimulate further studies, especially microscopic, to elucidate those questions.

%
%
%
%
%

\section{Conclusion}
\label{Conclusion}

To conclude, we have performed a detailed ac susceptibility study of Cu$_2$OSeO$_3$ in and around the SkL phase. The dissipation revealed through the imaginary component of the susceptibility shows a complex evolution with magnetic field. A narrow peak that appears for small magnetic fields at the helical-to-paramagnet transition grows in magnitude up to the SkL pocket where it starts to move away from the paramagnetic phase, marking the boundary between the SkL and the conical phase. A second peak appears that follows the SkL-to-paramagnet transition, with similar frequency and magnetic field dependence. We have successfully modelled the frequency dependence of the dissipation inside the SkL phase in terms of a distribution of relaxation times. The characteristic frequency spans several orders of magnitude in a narrow temperature range where the SkL phase exists, indicating that the source of dissipation is linked to SkL.

%
%
%
%
%

\section{Acknowledgement}
\label{Acknowledgement}

The support from the Croatian Science Foundation Project No. 02.05/33 and the Swiss NCCR and its programme MaNEP are acknowledged.

\bibliography{Cu2OSeO3}

\end{document}